\documentclass[sigconf]{acmart}
\usepackage{xspace}
\usepackage[textsize=tiny]{todonotes}
\usepackage{bbding}
\usepackage{pifont}
\usepackage[english]{babel}

\usepackage{threeparttable}
\usepackage{multirow}
\usepackage{subcaption}
\usepackage{amsmath}
\newcommand{\eat}[1]{}
\newcommand{\zhibo}[1]{\textcolor{blue}{#1}}  
\newcommand{\hansheng}[1]{}  
\usepackage{balance}
\newcommand{\ourmeth}{DUIN\xspace}

\usepackage{amsmath,amsfonts,bm}

\def\eqref#1{equation~\ref{#1}}

\def\1{\bm{1}}

\DeclareMathAlphabet{\mathsfit}{\encodingdefault}{\sfdefault}{m}{sl}
\SetMathAlphabet{\mathsfit}{bold}{\encodingdefault}{\sfdefault}{bx}{n}

\DeclareMathOperator*{\argmin}{arg\,min}

\AtBeginDocument{%
  }

\setcopyright{acmcopyright}
\copyrightyear{2018}
\acmYear{2018}
\acmDOI{XXXXXXX.XXXXXXX}

\acmConference[Conference acronym 'XX]{Make sure to enter the correct
  conference title from your rights confirmation emai}{June 03--05,
  2018}{Woodstock, NY}
\acmPrice{15.00}
\acmISBN{978-1-4503-XXXX-X/18/06}

\begin{document}
\title{Modeling User Intent Beyond Trigger: Incorporating Uncertainty for Trigger-Induced Recommendation}

\author{Jianxing Ma}
\email{majianxing.mjx@alibaba-inc.com}
\affiliation{%
  \institution{Alibaba Group}
  \city{Hangzhou}
  \state{Zhejiang}
  \country{China}
}

\author{Zhibo Xiao}
\email{xiaozhibo.xzb@alibaba-inc.com}
\affiliation{%
  \institution{Alibaba Group}
  \city{Hangzhou}
  \state{Zhejiang}
  \country{China}
}

\author{Luwei Yang}
\authornote{Corresponding author.}
\email{luwei.ylw@alibaba-inc.com}
\affiliation{
  \institution{Alibaba Group}
  \city{Hangzhou}
  \state{Zhejiang}
  \country{China}
}

\author{Hansheng Xue}
\email{hansheng.xue@nus.edu.sg}
\affiliation{
  \institution{National University of Singapore}
  \country{Singapore}
}

\author{Xuanzhou Liu}
\email{liuxuanzhou.lxz@alibaba-inc.com}
\affiliation{%
  \institution{Alibaba Group}
  \streetaddress{}
  \city{Hangzhou}
  \state{Zhejiang}
  \country{China}
}

\author{Wen Jiang}
\email{wen.jiangw@alibaba-inc.com}
\affiliation{%
  \institution{Alibaba Group}
  \streetaddress{}
  \city{Hangzhou}
  \state{Zhejiang}
  \country{China}
}

\author{Wei Ning}
\email{wei.ningw@alibaba-inc.com}
\affiliation{%
  \institution{Alibaba Group}
  \streetaddress{}
  \city{Hangzhou}
  \state{Zhejiang}
  \country{China}
}

\author{Guannan Zhang}
\email{zgn138592@alibaba-inc.com}
\affiliation{%
  \institution{Alibaba Group}
  \streetaddress{}
  \city{Hangzhou}
  \state{Zhejiang}
  \country{China}
}
\renewcommand{\shortauthors}{Jianxing Ma et al.}
\begin{abstract}
     \eat{Numerous e-commerce platforms offer novel recommendation scenarios to meet user demands for an immersive browsing experience. In these scenarios, the recommendations are constrained by the clicked trigger item, giving rise to a new recommendation task referred to as Trigger-Induced Recommendation (TIR). Existing TIR methods predominantly depend on the trigger item to grasp user intent. However, they lack higher-level exploration and exploitation of user intent (e.g., popular items and complementary items), potentially resulting in an overly convergent understanding of user intent, which can be detrimental to user's long-term purchasing experiences. Moreover, user intent exhibits uncertainty and is affected by many factors such as browsing context and historical behaviors, posing a challenge in modeling user intent. To address these challenges, we propose a novel model called \textbf{D}eep \textbf{U}ncertainty \textbf{I}ntent \textbf{N}etwork (DUIN~\footnote{The source code of \ourmeth will be released at: https://github.com/xxx.}), comprising three essential modules: 1) The Explicit Intent Exploit Module (EIEM) is introduced to extract explicit user intent using the contrastive learning paradigm; 2) The Latent Intent Explore Module (LIEM) is designed to explore latent user intent by leveraging the multi-view relationships between items, applied to both the trigger item and the target item to uncover user latent intent; 3) The Intent Uncertainty Measurement Module (IUMM) is proposed to provide a distributional estimation and capture the uncertainty associated with user intent. 
     Experiments on three real-world datasets demonstrate the superior performance of DUIN compared to existing baselines.
     Significantly, DUIN has been implemented in all TIR scenarios of our e-commerce application, and the outcomes from online A/B testing conclusively verify the superiority of our approach.}
     \eat{To cater to users' desire for an immersive browsing experience, numerous e-commerce platforms provide various recommendation scenarios, with a focus on Trigger-Induced Recommendation (TIR) tasks. 
     However, the majority of current TIR methods heavily rely on the trigger item to understand user intent, lacking a higher-level exploration and exploitation of user intent (e.g., popular items and complementary items), which may result in an overly convergent understanding of users' short-term intent and harm users' long-term purchasing experiences. 
     Moreover, users' short-term intent show uncertainty and are affected by various factors such as browsing context and historical behaviors, \zhibo{ posing challenges to user intention modeling}.
     To address these challenges, we propose a novel model called \textbf{D}eep \textbf{U}ncertainty \textbf{I}ntent \textbf{N}etwork (DUIN), comprising three essential modules: 1) The Explicit Intent Exploit Module (EIEM) is introduced to extract explicit user intent using the contrastive learning paradigm; 2) The Latent Intent Explore Module (LIEM) is designed to explore latent user intent by leveraging the multi-view relationships between items, applied to both the trigger item and the target item to uncover user latent intent; 3) The Intent Uncertainty Measurement Module (IUMM) is proposed to provide a distributional estimation and capture the uncertainty associated with user intent. 
     Experiments on three real-world datasets demonstrate the superior performance of DUIN compared to existing baselines. Significantly, DUIN has been implemented in all TIR scenarios of our e-commerce application, and the outcomes from online A/B testing conclusively verify the superiority of our approach.}
     To cater to users' desire for an immersive browsing experience, numerous e-commerce platforms provide various recommendation scenarios, with a focus on Trigger-Induced Recommendation (TIR) tasks. However, the majority of current TIR methods heavily rely on the trigger item to understand user intent, lacking a higher-level exploration and exploitation of user intent (e.g., popular items and complementary items), which may result in an overly convergent understanding of users' short-term intent and can be detrimental to users' long-term purchasing experiences. Moreover, users' short-term intent shows uncertainty and is affected by various factors such as browsing context and historical behaviors, which poses challenges to user intent modeling.
     To address these challenges, we propose a novel model called \textbf{D}eep \textbf{U}ncertainty \textbf{I}ntent \textbf{N}etwork (\textbf{DUIN}), comprising three essential modules: 
     i) Explicit Intent Exploit Module extracting explicit user intent using the contrastive learning paradigm; ii) Latent Intent Explore Module exploring latent user intent by leveraging the multi-view relationships between items;
     iii) Intent Uncertainty Measurement Module offering a distributional estimation and capturing the uncertainty associated with user intent. 
     Experiments on three real-world datasets demonstrate the superior performance of DUIN compared to existing baselines. Notably, DUIN has been deployed across all TIR scenarios in our e-commerce platform, with online A/B testing results conclusively validating its superiority.
\end{abstract}

\ccsdesc[500]{Information systems~Personalization}
\ccsdesc[500]{Information systems~Recommender systems}
\ccsdesc[300]{Information systems~Learning to rank}

\keywords{Click-Through Rate Prediction, Trigger-Induced Recommendation, Contrastive Learning, User Intent Modeling}

\maketitle

\begin{figure}[tb]
\vspace{-0.5em}
  \centering
 \setlength{\abovecaptionskip}{0.2cm} \includegraphics[width=0.9\columnwidth,height=0.3\textwidth]{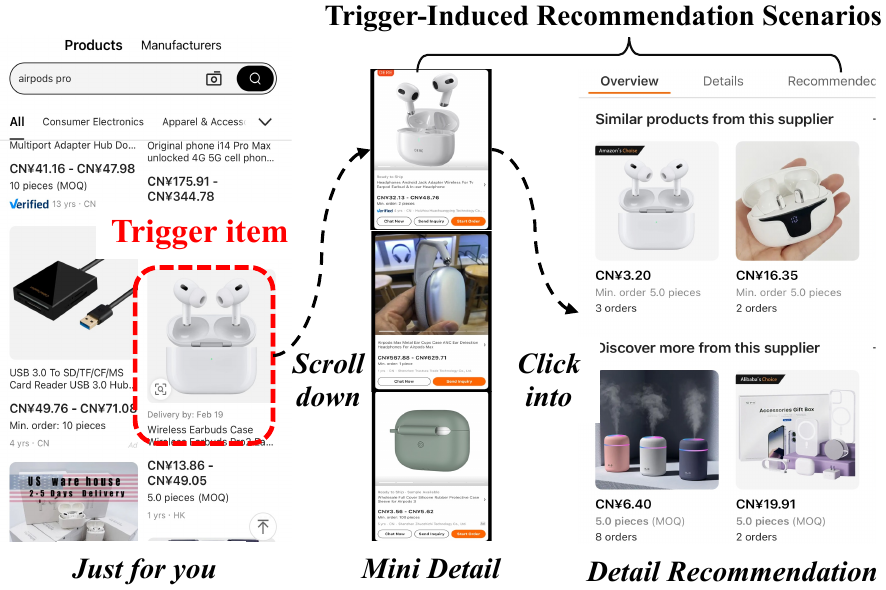}
  \caption{Recommendation scenarios of an e-commerce platform. Left: Just for You. Middle: Mini Detail. Right: Detail Recommendation.}
  \label{fig:recom_demo}
  \setlength{\belowcaptionskip}{0.5cm}
  \vspace{-2.0em}
\end{figure}
\section{Introduction}

Trigger-Induced Recommendation (TIR)~\cite {Shen2022DIHN,Xia2023DIAN,xiao2024deep} has recently attracted significant industry attention for its ability to deliver immersive user experiences by modeling the instant history of user clicks on trigger items. For instance, as one of the world's largest Business-to-business (B2B) e-commerce platforms, \textit{Alibaba.com~\footnote{https://www.alibaba.com/}} also provides users with many TIR scenarios. As illustrated in Figure~\ref{fig:recom_demo}, by clicking a trigger item in \textit{Just for You} scenario, users can access some TIR scenarios, e.g., \textit{Mini Detail} or \textit{Detail Recommendation}. 
The primary objective of TIR is to recommend items satisfying users' needs on conditional of the instant clicking of the trigger item. Nowadays, TIR is playing an increasingly significant role in many industries~\cite{Shen2022DIHN,Xie2021R3S}. More than half of active buyers on our website are contributed by TIR among all recommendation scenarios.

\begin{figure}[tb]
    \centering
    \begin{subfigure}{4cm}
        \includegraphics[width=4.3cm]{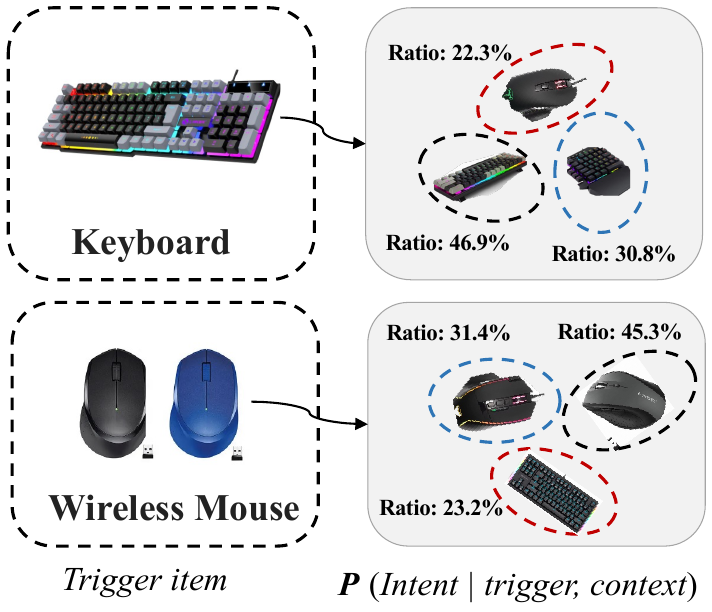}
        \caption{\ourmeth}
        \label{fig:sub2}
    \end{subfigure}
    \hfill 
     \begin{subfigure}{4cm}
        \includegraphics[width=4.3cm]{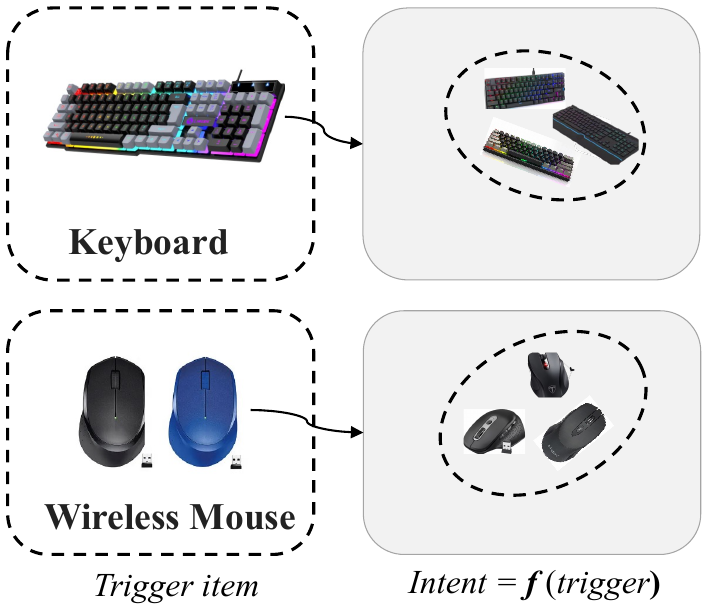}
        \caption{Existing TIR methods}
        \label{fig:sub1}
    \end{subfigure}
    \caption{Differences in user intent modeling between existing TIR methods and \ourmeth. The black dashed frame are similar items, red indicates complementary items, and blue denotes trending items. Ratio denotes the proportion of purchasing users. Best viewed in color.}
    \label{fig:duin_demo}
    \setlength{\belowcaptionskip}{-0.9cm}
    \vspace{-1.5em}
\end{figure}

Traditional recommendation modeling methods~\cite{Cheng2016WDL, Wang_Zhang_Xie_Guo_2018,Li_Cheng_Chen_Chen_Wang_2020,Lyu_Dong_Huo_Ren_2020,Pi_Zhou_Zhang_Wang_Ren_Fan_Zhu_Gai_2020} often focus on modeling user chronological behaviors with various deep neural networks~\cite{sun2019bert4rec,Liu_Tang_Chen_Yu_Guo_Zhang_2019,xu2019recurrent}, lacking consideration for the trigger item, and thus cannot be directly applied to the TIR scenario. Recently, several trigger-based methods~\cite{Shen2022DIHN,Xie2021R3S,xiao2024deep} have been proposed for the TIR task, which have demonstrated their superiority over traditional methods in the TIR scenario. Nonetheless, user behaviors in real-world e-commerce platforms are more complex~\cite{Xu_He_Tan_Li_Lang_Guo_2020,Qin_Zhang_Wu_Jin_Fang_Yu_2020,Li_Wang_Tan_Zeng_Ou_Ou_Zheng_2020,wang2021dcn}. Figure~\ref{fig:sub2} shows an insightful example of users' subsequent behaviors after clicking a trigger item.

When users clicked on a keyboard, 
$46.9\%$ of users subsequently purchased the same or similar items (e.g., another keyboard), while $30.8\%$ of them bought trending products (e.g., wireless keyboard) and the rest $22.3\%$ of them bought complementary options (e.g., wireless mouse). This observation reveals that user intent after clicking on a trigger item exhibits a certain degree of uncertainty. 
This also indicates a need for the platform to offer more precise and comprehensive procurement recommendations.

However, as shown in Figure~\ref{fig:sub1}, existing TIR methods overly rely on the trigger item, neglecting deeper exploration and exploitation of user intent, which results in an excessively convergent understanding of user intent, and the user will be isolated in a small set of recommended items from the trigger item. For users purchasing computer accessories, products like keyboards and wireless mice are within their intended scope, the overly narrow recommendations can hinder the optimization of users' long-term purchasing experience. Furthermore, user intent is shaped by a multitude of factors, such as user historical interactions and browsing context. This complexity presents a significant challenge in accurately modeling user intent and capturing the associated uncertainty.

To address these challenges, we propose a novel model in this paper called \textbf{\ourmeth~\footnote{The source code of \ourmeth will be released at: https://github.com/majx1997/DUIN.}}, short for \textbf{D}eep \textbf{U}ncertainty \textbf{I}ntent \textbf{N}etwork. It primarily comprises three components. 
To extract explicit user intent based on the trigger, we design an \textit{Explicit Intent Exploit Module} (EIEM) that uses the contrastive learning paradigm to obtain generalizable and distinguishable intent representations. 
We also design a \textit{Latent Intent Explore Module} (LIEM) to effectively explore the users' underlying intent by leveraging the multi-view relationships between items, applicable to both the trigger item and the target item. 
An \textit{Intent Uncertainty Measurement Module} (IUMM) is implemented to model user intent intensity as a Gaussian distribution, capturing the uncertainty associated with user intent. 
Overall, the contributions of this paper can be summarized as follows:

\begin{itemize}
    \item We propose a novel model, Deep Uncertainty Intent Network (\ourmeth), to address the uncertainty challenges of user intent modeling in Trigger-Induced Recommendation.
    \item We introduce an Explicit Intent Exploit Module to extract explicit user intent, and also a Latent Intent Explore Module to explore latent user intent.
    \item We design a new Intent Uncertainty Measurement Module to offer distribution estimates and capture the uncertainty of user intent.
    \item Extensive experiments on three real-world datasets and an industrial platform demonstrate the superior performance of our model compared to state-of-the-art baselines.
\end{itemize}

\section{Related Work}

\textbf{User behavior Modeling}. 
Existing user behavior modeling approaches~\cite{Zhou2018DIN,Song_Huang_Zhang_Lu_2021,Gharibshah_Zhu_Hainline_Conway_2020,pi2019practice} primarily focus on learning better representation from user behavior sequences and extracting the users' interests for personalized recommendation. 
In particular, attention mechanisms\cite{wang2019kgat,tay2018multi,li2020time,song2019session,xu2019graph} have gained popularity in recent years due to their ability to capture complex patterns and temporal dynamics manifested within behavior sequences. For example, 
DIN~\cite{Zhou2018DIN} utilizes attention mechanisms to activate the interest relation between historical behaviors and the target item. DIEN~\cite{Zhou2019DIEN} further integrates a special attention structure with a GRU module to capture the dynamic evolution of user interests. DSIN~\cite{Feng2019DSIN} divides user's sequential behaviors into multiple sessions and employs a self-attention layer to model inner-session interests, alongside a bi-directional LSTM to capture intra-session interests. DMIN~\cite{Xiao2020DMIN} explores user’s multiple diverse interests and introduces a multi-head self-attention layer and a multi-interest extractor layer to represent user’s interests by multiple vectors.
These attention-based methods have demonstrated feasible performance across various scenarios.
Although the aforementioned recommendation approaches can be directly applied to TIR scenarios, their inability to account for modeling user's immediate intent representing by the trigger item leads to challenges in accurately estimating user intent in real-world industrial scenarios.

\textbf{Trigger-Induced Recommendation}. Though above traditional solutions can be rigidly applied to TIR scenarios, the lack of considering the trigger item usually make them get suboptiomal results in these scenarios.
Recently, various methods have been proposed to target for the TIR problem. 
DIHN~\cite{Shen2022DIHN} is one pioneer in introducing the TIR task, and utilizes a deep network to predict the user's real intent regarding the trigger item. DIAN~\cite{Xia2023DIAN} proposes an intention-aware network to extract the user’s intention and balance the outcomes of trigger-free and trigger-based recommendations. DEI2N~\cite{xiao2024deep} introduces a user instant interest modeling layer to forecast the dynamic change in the intensity of instant interest when the user clicks on a trigger item and scrolls down.  
These existing works have showcased their superiority over traditional methods in the TIR scenario. 
However, the heavy reliance of these recently proposed methods on trigger items limits their capacity to capture the intricate and higher-level relationships in user behavior. As a result, they may fail to fully reveal the extent of user's true intent, potentially leading to an overly convergent understanding of user intent. This limitation could adversely affect users' long-term experiences.

\section{Method}
\begin{figure*}[th]
\setlength{\abovecaptionskip}{0.1cm}
   \centering
   \includegraphics[width=0.85\textwidth,height=0.425\textwidth]{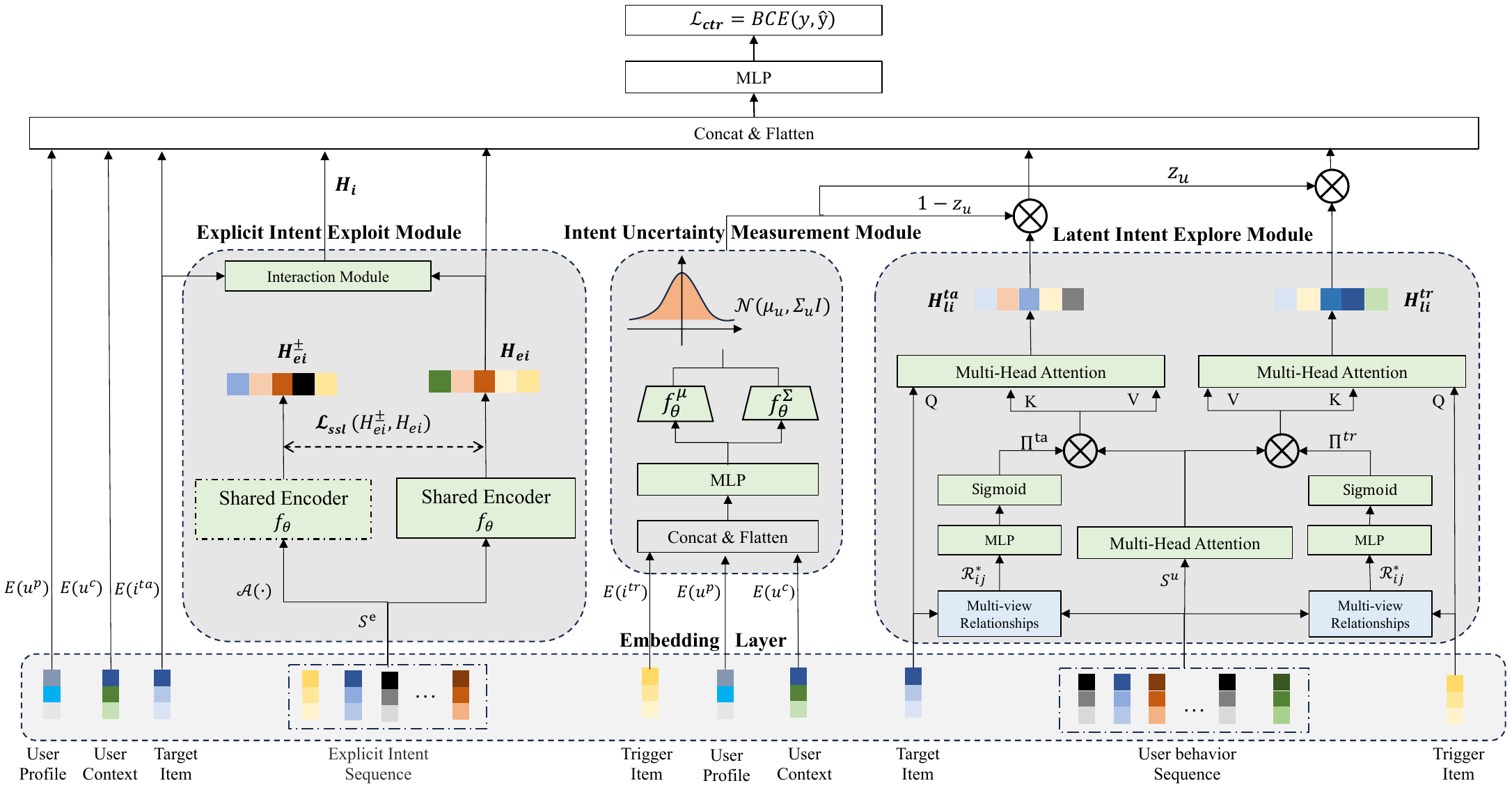}
     \caption{The architecture of \ourmeth consists of three modules, EIEM, LIEM and IUMM.}
      \label{fig:main_pic}
      \vspace{-0.5em}
 \end{figure*}
\eat{The overall architecture of the proposed \ourmeth is illustrated in Figure~\ref{fig:main_pic}, consisting primarily of three modules: 1) Explicit Intent Exploit Module (EIEM), which employs  contrast learning framework to extract user explicit intent based on trigger item; 2) Latent Intent Explore Module (LIEM), which is responsible to explore user latent intent based on multiple co-occurrence views; 3) Intent Uncertainty Measurement Module (IUMM), which gives a distributional estimation in the latent space and inject uncertainty into the intent modeling.}

The overall architecture of the proposed \ourmeth is illustrated in Figure~\ref{fig:main_pic}, which mainly consists of three modules: Explicit Intent Exploit Module (EIEM), Latent Intent Explore Module (LIEM), and Intent Uncertainty Measurement Module (IUMM).
Before diving into the details of these modules, we first formulate the TIR task and introduce some necessary notations.

\subsection{Problem Setup}
Let $\mathcal{U}=\{u_1,u_2,...,u_M\}$ and $\mathcal{I}=\{i_1,i_2,...,i_N\}$ represent the user set and the item set, respectively. Each user has a behavior sequence $\mathcal{S}^u=\{i_1^u,i_2^u,...,i_T^u\}$, where $M$, $N$, and $T$ denote the number of users, items, and the length of behavior sequences, respectively. We denote user profile features as $u^p$ and context features as $u^c$. Based on these notations, given the trigger item $i^{tr}$, the primary objective of TIR is to predict the probability $\hat{y}$ of user $u$ potentially interacting with the target item $i^{ta}$ as follows:
\begin{equation}
    \hat{y}=\mathcal{F}_\theta(\mathbf{E}(u^p),\mathbf{E}(u^c),\mathbf{E}(i^{tr}),\mathbf{E}(i^{ta}),\mathbf{E}(\mathcal{S}^u)),
    \label{eq:ctr}
\end{equation}
where $\mathcal{F}_\theta$ represents the prediction function we aim to learn. The notation $\mathbf{E}(\cdot)\in \mathbb{R}^d$ denotes the embedding layer~\cite{Cheng2016WDL}, which transforms raw features into dense vectors with dimension $d$.

\subsection{Deep Uncertainty Intent Network}
\eat{In this section, we present the Deep Uncertainty Intent Network (\ourmeth). The overall architecture of \ourmeth is illustrated in Figure~\ref{fig:main_pic}.
The overall architecture of the proposed \ourmeth is illustrated in Figure~\ref{fig:main_pic}, consisting primarily of three modules: explicit intent exploit, latent intent explore, and intent uncertainty measurement.}
\subsubsection{\textbf{Explicit Intent Exploit Module}} 
User chronological behaviors exhibit causality and convey explicit intent information, including aspects users are inclined to interact with, such as brand, theme and category. The trigger item merely represents a partial expression of user explicit intent. In \ourmeth, we introduce the Explicit Intent Exploit Module to learn user explicit intent.

Typically, items with identical attributes exhibit a high degree of similarity. For instance, in the e-commerce scenario, products that belong to the same category often share similar intrinsic quality or interacted uses. In user behavior sequences, we consider items that share attributes with the trigger item express the user explicit intent.\eat{We consider that items in user historical behaviors, which have the same attribute as the trigger item, express the user explicit intent to purchase.} Given a trigger item $i^{tr}$, we retrieve items from the behavior sequence $\mathcal{S}^u$ that have the same attribute as $i^{tr}$. Together with the trigger item, these form an explicit intent sequence $\mathcal{S}^e$: 
\begin{equation}
    \mathcal{S}^e=\{i^{tr},i_1^e,...i_l^e,...,i_L^e\}, 
\end{equation}
where $i_l^e$ denotes items that share the same attribute with the trigger item $i^{tr}$. $L$ represents the number of $i_l^e$.

 Various attention-based methods~\cite{Zhou2018DIN,Kang_McAuley_2018} have significantly facilitated user behavior modeling. However, recent research~\cite{Qiu_Huang_Yin_Wang_2022,Xie_Sun_Liu_Wu_Gao_Zhang_Ding_Cui_2022,chen2022intent} indicates that user representations generated by these methods tend to degenerate into an anisotropic shape, which may result in high semantic similarities among representations. To address this issue, inspired by recent advances of contrastive learning paradigm in the computer vision~\cite{Chen_Xie_He_2021,Chen_Kornblith_Norouzi_Hinton_2020}, based on the explicit intent sequence, we apply contrastive learning to obtain discriminative and high-quality user explicit intent representations. Formally, the explicit intent contrastive learning task can be defined as: 
\begin{equation}
    f_{\theta} = \argmin\limits_{f_\theta}\mathcal{L}_{ssl}(f_\theta(\mathcal{S}^e), f_\theta(\widetilde{\mathcal{S}}^e_{+})),
\end{equation}
where $\mathcal{L}_{ssl}$ is the contrastive loss function. $f_\theta$ is the shared encoder. $\widetilde{\mathcal{S}}^e_{+}$ is the augmented view of the explicit intent sequence $\mathcal{S}^e$.

In practice, we mask the explicit intent sequence $\mathcal{S}^e$ with a certain probability to get the augmented view:
\begin{equation}
    \widetilde{\mathcal{S}}^e_{+}\sim \mathcal{A}(\mathcal{S}^e, \gamma), 
\end{equation}
where $\mathcal{A}(\cdot)$ denotes the augmentation operator and $\gamma$ is the mask probability. The idea behind this is that user explicit intent is relatively stable during a period of time. Therefore, though some of the items are masked, the explicit intent information is still retained and should be considered as the positive view. Thus, we treat $(\mathcal{S}^e, \widetilde{\mathcal{S}}^e_{+})$ as positive view pairs and other augmented examples within the same batch as negative views $\widetilde{\mathcal{S}}^e_{-}$.

Furthermore, we adopt the shared encoder $f_\theta$\eat{and a separate projector head $p_\phi$} to extract the user explicit intent representation $\mathcal{H}_{ei} = f_\theta (S^e)$. We optimize $f_\theta$ via a contrastive loss function~\cite{Chen_Kornblith_Norouzi_Hinton_2020}, which can be formulated as: 
\begin{equation}
    \mathcal{L}_{ssl} = -log\frac{e^{sim(\mathcal{H}_{ei},\mathcal{H}_{ei}^{+})/\tau}}{e^{sim(\mathcal{H}_{ei},\mathcal{H}_{ei}^{+})/\tau}+\sum_{i=1}^{2B-1}e^{sim(\mathcal{H}_{ei},\mathcal{H}_{ei}^{-})/\tau}},
\end{equation}
where $\mathcal{H}_{ei}^{+}$ and $\mathcal{H}_{ei}^{-}$ indicate the augmented positive view and the negative views of $\mathcal{H}_{ei}$ respectively. $sim(\cdot)$ denotes the cosine similarity function. $B$ is the batch size. $\tau$ is the temperature parameter.

It is worth mentioning that EIEM also introduces a feature interaction module to learn the explicit interaction relationship between the target item $i^{ta}$ and the user explicit intent $\mathcal{H}_{ei}$, the output representation $\mathcal{H}_{i}$ is calculated as:
\begin{equation}
    \mathcal{H}_{i} = MLP(\mathcal{H}_{ei},\mathbf{E}(i^{ta}),\mathcal{H}_{ei}\odot\mathbf{E}(i^{ta}),\mathcal{H}_{ei}-\mathbf{E}(i^{ta})),
\end{equation}
where $\odot$ denotes the Hadamard product.
\subsubsection{\textbf{Latent Intent Explore Module}} As aforementioned, users have various latent intents after clicking on the trigger item, such as seeking similar items, popular items, or complementary items. Previous TIR methods~\cite{Shen2022DIHN,Xia2023DIAN,xiao2024deep} lack exploration of user latent intent and overlook the potential information related to user intent within user behavior sequences. In \ourmeth, we propose the Latent Intent Explore Module to reveal user latent intent. 

 We consider users with similar intents usually exhibit similar behaviors, thus, we aggregate the temporal directed behaviors of multiple users to form a fundamental probabilistic graph structure $\mathcal{G}$. The nodes $\mathcal{V}$ in the graph contain items and their attributes, and we will connect two sequentially occurring nodes to refine collaborative relationships~\cite{wu2019session,chang2021sequential,min2022neighbour}. 
 
 Having obtained the finer-grained collaborative patterns in graph $\mathcal{G}$, suppose there is an item $i$ with attribute $\phi$ and an item $j$ with attribute $\psi$, we construct multi-view relationships between item $i$ and item $j$ as follows:
 \begin{align}
    \mathcal{R}_{ij}^{t} &= \mathcal{C}_{i \rightarrow j}(\{(\mathcal{V}_i, \mathcal{V}_j)|(\mathcal{V}_i,\mathcal{V}_j)\in\mathcal{G}\}), \\
    \mathcal{R}_{ij}^{c} &= \mathcal{C}_{\phi \rightarrow \psi}(\{(\mathcal{V}_\phi, \mathcal{V}_\psi)|(\mathcal{V}_\phi,\mathcal{V}_\psi)\in\mathcal{G}\}), \\
    \mathcal{R}_{ij}^{p} &= \mathcal{C}_{\phi \rightarrow j}(\{(\mathcal{V}_\phi, \mathcal{V}_j)|(\mathcal{V}_\phi,\mathcal{V}_j)\in\mathcal{G}\}),
\end{align}
where $\mathcal{V_*}$ denotes the note in graph $\mathcal{G}$. $\mathcal{C_*}$ is the aggregation operator that records two nodes' co-occurrence frequency. 

The high-order connections between items can be implicitly captured by the aforementioned multi-view relationships. Specifically, we use $\mathcal{C}_{i \rightarrow j}$ to calculate the co-occurrence frequency between item $i$ and item $j$ and it is straightforward to use $\mathcal{R}_{ij}^{t}$ to represent their directed transition relationship. Similarly, the collaborative relationship of attributes, denoted by $\mathcal{R}_{ij}^{c}$, represents the complementary relationship between item $i$ and item $j$. $\mathcal{R}_{ij}^{p}$ is utilized to characterize the popularity of item $j$ under attribute $\phi$. 

For item $i$ and item $j$, to better encode the information from multi-view relationships between them, we define the latent intent relevance score $\Pi_{ij}$ to quantify their potential connections\eat{as follows}:
\begin{equation}
    \Pi_{ij} = Sigmoid(MLP(\mathbf{E}(\mathcal{R}_{ij}^{t}), \mathbf{E}(\mathcal{R}_{ij}^{p}), \mathbf{E}(\mathcal{R}_{ij}^{c}))).
    \label{eq:intent_score}
\end{equation}

Motivated by previous work~\cite{Shen2022DIHN,xiao2024deep}, we employ Multi-Head Self-Attention (MHSA)~\cite{Vaswani_Shazeer_Parmar_Uszkoreit_Jones_Gomez_Kaiser_Polosukhin_2017} to obtain the refined behavior representation $\mathbf{E}(\mathcal{S}^u)$. Next, we apply two MHSA modules to extract user latent intent with respect to the target item and the trigger item respectively. More importantly, we propose a simple yet effective method to enable that the attention mechanism aware of the prior multi-view relationships among items. Specifically, for each behavior item $i_*^u\in\mathcal{S}^u$, we separately calculate the latent intent relevance score with both the trigger item and the target item according to \eqref{eq:intent_score}, and we also modify the attention calculation scheme as follows:
\begin{align}
    \mathcal{H}_{li}^{tr} &= Attention(\mathbf{E}(i^{tr}),\Pi^{tr}\mathbf{E}(\mathcal{S}^u),\Pi^{tr}\mathbf{E}(\mathcal{S}^u)), \\
    \mathcal{H}_{li}^{ta} &= Attention(\mathbf{E}(i^{ta}),\Pi^{ta}\mathbf{E}(\mathcal{S}^u),\Pi^{ta}\mathbf{E}(\mathcal{S}^u)),
\end{align}
where $\mathcal{H}_{li}^{tr}$ and $\mathcal{H}_{li}^{ta}$ are user latent intent representation with respect to the trigger item and the target item respectively. $\Pi^{tr}$ represents the latent intent relevance score between each behavior item and the trigger item. $\Pi^{ta}$ denotes the relevance score between each behavior item and the target item. 

\subsubsection{\textbf{Intent Uncertainty Measurement Module}}In the TIR task, supposing a continuous mapping space $\mathcal{X}\rightarrow\mathcal{Z}$, where $x_u\in\mathcal{X}$ denotes user personalized information, including the trigger item $i^{tr}$, context features $u^c$, and profile features $u^p$. $z_u\in\mathcal{Z}$ represents user intent intensity. As user intent evolves and browsing context changes, the intensity of user intent after clicking on the trigger item becomes uncertain to some degree. Therefore, in \ourmeth, we consider the user interaction behavior as an uncertain event influenced by the user intent intensity. We employ the Intent Uncertainty Measurement Model to model user intent intensity as a distribution rather than a static value~\cite{Shen2022DIHN,xiao2024deep}, which can cover a broader space of user intent and infuse the recommendation results with novelty and uncertainty.

Specifically, we define the user intent intensity $z_u$ of each user as a Gaussian distribution $\mathcal{N}$, which is mathematically defined as:
\begin{equation}
    \mathit{p}(z_u|x_u) = \mathcal{N}(z_u;\mu_u, \Sigma_u\textbf{\textit{I}}), 
\end{equation}
where $\mu_u$ and $\Sigma_u$ denote the mean and the variance of the Gaussian distribution, respectively. $\mu_u$ represents the predicted user intent intensity and $\Sigma_u$ can be regarded as the uncertainty of $\mu_u$. Here we only consider a diagonal matrix for simplicity. In practice, we employ two independent networks to obtain $\mu_u$ and $\Sigma_u$:
\begin{align}
    \mu_u &= f_\theta^\mu(x_u), \\
    \Sigma_u &= softplus(f_\theta^\Sigma(x_u)),
\end{align}
where $f_\theta^\mu$ and $f_\theta^\Sigma$ are implemented as two fully-connected layers. $softplus$ is an activation function~\cite{glorot2011deep} and constrains the output to always be positive.

Finally, we consider $z_u$ as a measure of the user's intent intensity towards the trigger item. We use it to integrate the user latent intent features outputted by LIEM, which can be formulated as:

\begin{equation}
    \mathcal{H}_{li} = (z_u\odot\mathcal{H}_{li}^{tr};(1-z_u)\odot\mathcal{H}_{li}^{ta}).
\end{equation}
\subsection{Prediction and Optimization}
Having obtained the user intent representations $\mathcal{H}_{ei}$, $\mathcal{H}_{i}$ and $\mathcal{H}_{li}$ from both EIEM and LIEM, similar to \eqref{eq:ctr}, we concatenate these intent representations with other features, including context features $u^c$, profile features $u^p$, and the target item $i^{ta}$. Subsequently, the concatenated features are fed into MLP~\cite{Zhou2018DIN} to estimate the likelihood $\hat{y_i}\in[0,1]$ of user interaction. We adopt the binary cross entropy loss~\cite{Zhou2018DIN,Shen2022DIHN,Wang2017DCN} for model optimization: 
\begin{equation}
    \mathcal{L}_{ctr} = -\frac{1}{N}\sum\limits_{i=1}^{N}(y_ilog(\hat{y_i}) + (1-y_i)log(1-\hat{y_i})),
\end{equation}
where $N$ denotes total size of the training set, $y_i$ is the ground-truth.
We also adopt a multi-task learning strategy to jointly optimize the original prediction task and the contrastive learning objectives. The final optimization function is defined as:
\begin{equation}
  \mathcal{L}_{final} = \mathcal{L}_{ctr} + \alpha\cdot\mathcal{L}_{ssl}, 
\end{equation}
where $\alpha$ is a hyperparameter to balance weight of contrastive loss.

\section{Evaluation}
\subsection{Experimental Setup}
\subsubsection{Datasets.} Three real-world datasets are used in experiments. Detailed statistics are summarized in Table~\ref{tab:dataset}.

\begin{itemize}
    \item \textbf{Alibaba.com}~\footnote{https://www.alibaba.com/}. As there is no public dataset released for the TIR task. We collect interaction behavior logs from the TIR scenarios on our e-commerce application to serve as industrial production dataset, named Alibaba.com. The dataset contains about 370 thousand users and 5.2 million records of browsing and clicking behavior data over 3 days. We select data from the first 80\% in chronological order as the training set, the next 10\% as validation set, while the rest 10\% as test set.
    \item \textbf{Alimama}~\footnote{https://tianchi.aliyun.com/dataset/dataDetail?dataId=56}. It is a representative e-commerce dataset provided by Alimama advertising platform. Since it lacks logs of trigger items, following the data processing strategy by ~\cite{Shen2022DIHN}, for each sample, we take the latest item clicked by user within four hours as the trigger item. Finally, the dataset contains 8.55 million samples from half a million users.
    \item \textbf{ContentWise~}\cite{PerezMaurera2020ContentWise}. Different from Alibaba.com and Alimama datasets, the dataset is a popular multimedia recommendation dataset, which is collected from an Over-The-Top Media service and includes 2.5 million behavior data from approximately 26 thousand users. Similar to the Alimama dataset, we consider the latest clicked item by a user within eight hours as the trigger item. 
\end{itemize}
\begin{table}[tbh]
\vspace{-1.0em}
\setlength{\abovecaptionskip}{0.05cm}
  \caption{Datasets statistics.}
  \label{tab:dataset}
  \begin{tabular}{lcccc}
    \toprule
    Dataset &\#Users &\#Items &\#Attributes &\#Samples\\
    \midrule
    Alibaba.com & 373,852 & 4,715,150 & 6,736 & 5,200,000 \\
    Alimama & 500,000 & 846,812 & 12,978
    & 8,552,702\\
    ContentWise &26,186 & 1,268,988& 117,693 & 2,585,070 \\
  \bottomrule
  \end{tabular}
\setlength{\belowcaptionskip}{0.9cm}
\vspace{-1.0em}
\end{table}
\subsubsection{Compared Methods.} We include three groups of baseline methods for comparison. We firstly compare \ourmeth with some popular user behavior modeling methods: \textbf{WDL}~\cite{Cheng2016WDL}, \textbf{DIN}~\cite{Zhou2018DIN}, \textbf{DIEN}~\cite{Zhou2019DIEN}, \textbf{DMIN}~\cite{Xiao2020DMIN}. Furthermore, following~\cite{Shen2022DIHN, Xia2023DIAN, xiao2024deep}, we upgrade previous methods by incorporating the trigger item, forming the second group for a fair comparison:
\begin{itemize}
    \item \textbf{WDL$^{TIR}$} concatenates the trigger item with other features and feds them into the deep side of the network.
    \item \textbf{DIN$^{TIR}$} employs an attention mechanism to activate related user behaviors with respect to the trigger item.
    \item \textbf{DIEN$^{TIR}$} utilizes GRU with an attentional gate to extract the evolved interests that are related to the trigger item.
    \item \textbf{DMIN$^{TIR}$} applies interest extraction layers to obtain user multiple interests that are related to the trigger item.
\end{itemize}
The third group involves several state-of-the-art methods focusing on the TIR task: \textbf{DIHN}~\cite{Shen2022DIHN}, \textbf{DIAN}~\cite{Xia2023DIAN}, \textbf{DEI2N}~\cite{xiao2024deep}.

\subsubsection{Implementation details.} We employ the Adam optimizer for all models with a learning rate of 0.001 and a batch size of 256. Each model is trained from scratch without any pre-training. The maximum lengths of the user sequences in the Alibaba.com, Alimama, and ContentWise datasets are 20, 20, and 30, respectively. We set the maximum length of user explicit intent sequences to 10 for the Alibaba and Alimama datasets and to 20 for the ContentWise dataset. We select category and brand as item attributes for e-commerce dataset and multimedia dataset respectively. We construct probability graph by linking each item to its four subsequent items. We utilize the Transformer~\cite{Vaswani_Shazeer_Parmar_Uszkoreit_Jones_Gomez_Kaiser_Polosukhin_2017} architecture as the encoder $f_\theta$, the number of attention heads is set to 8 for all. 
The MLP within the EIEM and IUMM are configured with hidden layer sizes of 144 and 72. The dimensions of the final hidden layers are set at 200 and 80, respectively. For contrastive learning, the hyperparameters $\tau$, $\gamma$, and $\alpha$ are assigned values of 0.1, 0.5, and 1, respectively.

\subsubsection{Evaluation Metrics.} 
In offline experiments, we use commonly used metrics AUC and RelaImpr \cite{Zhou2019DIEN,Xiao2020DMIN,Feng2019DSIN}, which are widely recognized evaluation metrics in CTR tasks. We conduct each experiment five times and report the average results. All other experimental and parameter settings remain consistent with ~\cite{Shen2022DIHN,xiao2024deep} to ensure fair comparison. 
One-sided Wilcoxon rank-sum p-value is calculated
to quantify the significance of the improvement achieved by DUIN~\cite{gehan1965generalized}.
In online A/B testing, we use CTR and Conversion Rate (CVR) as online metrics to assess the efficacy of \ourmeth.
All reproducible materials of DUIN will be available after paper review.

\begin{table*}[tb]
\centering
\label{tab: Experimental Results}
\caption{Performance comparison of our method with competitors on three real-world datasets. Bold values indicate the best result in each column, while underlined values indicate the second best result.}
\label{tab:Experimental Results}
{
\begin{threeparttable}
\begin{tabular}{lcccccc}
\toprule
    \multirow{2}{*}{Model} &\multicolumn{2}{c}{Alibaba.com}  &\multicolumn{2}{c}{Alimama} &\multicolumn{2}{c}{ContentWise}\\
    \cline{2-7}
     &AUC &RelaImpr &AUC &RelaImpr &AUC &RelaImpr \\
    \midrule
    {WDL} 
    &{$0.6096\pm0.0019$} &{$-0.99\%$}
    &{$0.6062\pm0.0008$} &{$-7.97\%$}
    &{$0.9469\pm0.0003$} &{$-7.28\%$} \\
    {DIN} 
    &{$0.6042\pm0.0016$} & {$-5.87\%$} 
    &{$0.6154\pm0.0007$} & {$0.00\%$} 
    &{$0.9774\pm0.0002$} & {$-0.95\%$}\\
    {DIEN} 
    &{$0.6047\pm0.0025$} & {$-5.42\%$} 
    &{$0.6155\pm0.0005$} & {$0.09\%$} 
    &{$0.9779\pm0.0013$} & {$-0.85\%$}\\
    {DMIN} 
    &{$0.6107\pm0.0011$} & {$0.00\%$} 
    &{$0.6154\pm0.0002$} & {$0.00\%$} 
    &{$0.9820\pm0.0002$} & {$0.00\%$}\\
    \hline 
    {WDL$^{TIR}$}
    &{$0.7412\pm0.0014$} & {$111.89\%$} 
    &{$0.6075\pm0.0018$} & {$-6.84\%$} 
    &{$0.9713\pm0.0004$} & {$-2.22\%$}\\
    {DIN$^{TIR}$} 
    &{$0.7425\pm0.0021$} & {$119.06\%$} 
    &{$0.6155\pm0.0015$} & {$0.09\%$} 
    &{$0.9803\pm0.0019$} & {$-0.35\%$}\\
    {DIEN$^{TIR}$} 
    &{$0.7419\pm0.0019$} & {$118.52\%$}
    &{$0.6157\pm0.0004$} & {$0.26\%$}
    &{$0.9796\pm0.0015$} & {$-0.50\%$}\\
    {DMIN$^{TIR}$}
    &{$0.7454\pm0.0007$} & {$121.68\%$}
    &{$0.6157\pm0.0003$} & {$0.26\%$}
    &{$0.9822\pm0.0003$} & {$0.04\%$} \\
    \hline
    {DIHN} 
    &{$0.7462\pm0.0006$} & {$122.40\%$}
    &{$0.6166\pm0.0008$} & {$1.04\%$}
    &{$0.9786\pm0.0012$} & {$-0.75\%$}\\
    {DIAN} 
    &{$0.7480\pm0.0016$} & {$124.03\%$}
    &{$0.6168\pm0.0002$} & {$1.21\%$}
    &{$0.9764\pm0.0003$} & {$-1.16\%$} \\
    {DEI2N} 
    &\underline{$0.7671\pm0.0012$} & \underline{$141.28\%$}
    &\underline{$0.6180\pm0.0005$} & \underline{$2.25\%$}
    &\underline{$0.9840\pm0.0002$} &\underline{${0.41\%}$} \\
    \hline 
    \textbf{{\ourmeth}}
    &{$\textbf{0.7782}\pm\textbf{0.0014}$}\tnote{*} & {$\textbf{151.31\%}$}
    &{$\textbf{0.6194}\pm\textbf{0.0003}$}\tnote{*} & {$\textbf{3.47\%}$}
    &{$\textbf{0.9881}\pm\textbf{0.0004}$}\tnote{*} & {$\textbf{1.27\%}$} \\
  \bottomrule
  \end{tabular}
  \begin{tablenotes}
    \footnotesize\item[$\ast$]{Asterisks represent where \ourmeth's improvement over compared methods is significant (one-sided rank-num p-value <0.01).}
\end{tablenotes}
\end{threeparttable}
}
\end{table*}

\subsection{Offline Experiments}
\subsubsection{Overall Comparison}The overall performance of all methods across three datasets is summarized in Table~\ref{tab:Experimental Results}. From the results, we can observe that traditional CTR methods deliver the poorest results compared with other methods, indicating their inadequacy for the TIR task. By incorporating the trigger item, upgraded versions of these traditional methods demonstrate much improved performance. For example, DMIN$^{TIR}$ has a AUC gains $22.1\%$ over its original version for Alibaba.com dataset. It shows that an elaborative modeling of the trigger item is necessary.
Meanwhile, existing TIR methods, DIHAN, DIAN and DEI2N, employ carefully designed networks to further surpass the performance of previous upgraded traditional approaches. 
Notably, our proposed \ourmeth achieves the best performance across all three datasets. The results show that the AUC gains of our method \ourmeth over the existing strongest competitor DEI2N are $1.45\%$, $0.22\%$ and $0.41\%$ for Alibaba.com, Alimama and ContentWise datasets respectively. It validates the efficacy of considering explicit and latent intents and the corresponding uncertainties in TIR scenarios.
\begin{table}[tb]
\setlength{\abovecaptionskip}{0.15cm}
\caption{Ablation experimental results on Alibaba.com.} 
\label{tab:Ablation Results}
\resizebox{\columnwidth}{!}{
\centering
\begin{threeparttable}
  \begin{tabular}{cccccccc}
    \toprule
    \multirow{2}{*}{Model}&\multicolumn{5}{c}{Module}&\multicolumn{2}{c}{Alibaba.com}\\
    \cline{2-8}
     &EIEM& LIEM&IUMM&SSL&SII&AUC &RelaImpr\\
    \midrule
    1 &\ding{51}&\ding{51}&\ding{51}&\ding{51}&& 
    {$\textbf{0.7782}\pm\textbf{0.0014}$}&{$\textbf{0.0}\%$}\\
    2 &&\ding{51}&\ding{51}&&&
     {$0.7637\pm0.0011$}&{$-5.21\%$}\\
    3 &\ding{51}&&\ding{51}&\ding{51}&&
    {$0.7653\pm0.0013$}&{$-4.63\%$}\\
    4 &\ding{51}&\ding{51}&&\ding{51}&&
    {$0.7536\pm0.0009$}&{$-8.84\%$}\\
    5 &\ding{51}&\ding{51}&\ding{51}&&&
    {$0.7730\pm0.0015$}&{$-1.87\%$}\\
    6 &\ding{51}&\ding{51}&&\ding{51}&\ding{51}&
    {$0.7661\pm0.0007$}&{$-4.35\%$}\\
\bottomrule
\end{tabular}
\end{threeparttable}
}
\setlength{\belowcaptionskip}{-0.9cm}
\vspace{-1.5em}
\end{table}
\subsubsection{Ablation Study.} As Alibaba.com is collected from real-world TIR scenarios, we conduct several ablation experiments on it to further analyze the effectiveness of each module. Table~\ref{tab:Ablation Results} shows the experimental results, where SSL represents the contrastive learning framework in EIEM, and SII denotes that \ourmeth utilizes a static intent intensity to replace IUMM by following~\cite{Shen2022DIHN,xiao2024deep}. Compared to Model 2-4, Model 1 achieves the highest AUC results, indicating that all modules are effective for \ourmeth. The superior performance of Model 1 over Model 4 shows that modeling the user intent intensity is critical in the TIR task. The ablation results of Models 1 and 5 confirm that contrastive learning can help improve the performance of \ourmeth. By comparing {Models} 1 and 6, we can conclude that modeling user intent intensity as a distribution, rather than a static value is a more effective representation method.

\subsubsection{Hyper-parameters Analysis.} 
\begin{figure}[tbh]
\vspace{-1.0em}
\setlength{\abovecaptionskip}{0.06cm}
    \centering
    \begin{subfigure}{4.2cm}
        \includegraphics[width=4.2cm]{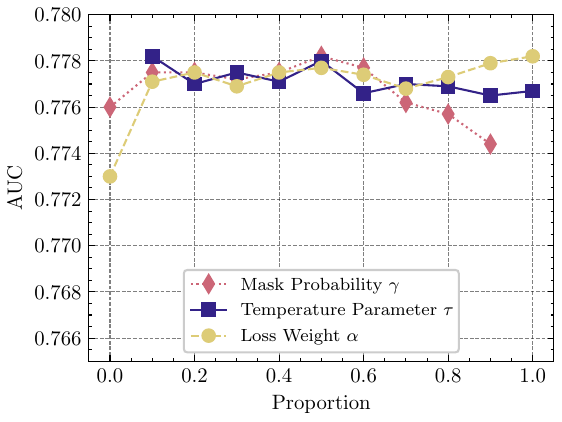}
        \caption{Hyper-parameters Analysis}
        \label{fig:analysis1}
    \end{subfigure}
    \hfill 
    \begin{subfigure}{4.2cm}
        \includegraphics[width=4.2cm]{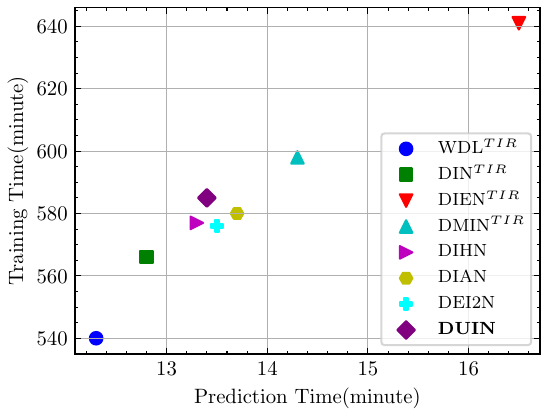}
        \caption{Time Analysis}
        \label{fig:analysis2}
    \end{subfigure}
    \caption{Parameters and Time Analysis on Alibaba.com.}
    \label{fig:analysis}
    \vspace{-1.5em}
\end{figure}

We further qualitatively analyze the impacts of different hyper-parameters in \ourmeth, which are shown in Figure \ref{fig:analysis1}. Overall, \ourmeth demonstrates relatively robust performance across parameter variations. We also observe that the performance of \ourmeth deteriorates when $\gamma$ is too large, probably because the excessive loss of interactive behaviors affects the modeling of user explicit intent.
\subsubsection{Time Analysis.} We also conduct an extra experiment to test the efficiency of \ourmeth. To make comparisons fairly, we conduct experiments on machines with identical hardware configurations, each of which has 1 NVIDIA Tesla V100 GPU with Intel(R) Xeon(R) Platinum 8163 CPU @ 2.50GHz and 330 GB memory. We report the time consumption for one epoch of training and testing for each model, respectively. As shown in Figure \ref{fig:analysis2}, compared to the existing TIR methods, \ourmeth slightly increases the training time. However, since contrastive learning is only utilized in the training stage, \ourmeth does not introduce additional prediction time cost.

\subsection{Online A/B Testing} 
We have implemented our proposed method, \ourmeth, in industrial Trigger-Induced Recommendation scenarios to serve substantial traffic requests. In this section, we will describe the online A/B testing settings and results.

\subsubsection{Online Serving Architecture} 
Figure \ref{fig:serving} depicts the online serving architecture of \ourmeth on our e-commerce platform. The architecture is primarily divided into two main stages: i) The offline stage focuses on training the DUIN model. In practice, we collect 60 days of user interaction logs from our platform and process these logs into training samples using the big data platform MAXCOMPUTE. Once the training samples are prepared, we utilize the Algorithm One Platform (AOP) to train our proposed \ourmeth model. After training is complete, the model is deployed for online serving. 
ii) During the online stage, The Personalization Platform (TPP) handles user requests by parsing them to obtain the real-time clicked trigger item and context features of the current browsing session. TPP then requests All Basic Feature Service (ABFS) to retrieve user profile features and historical behaviors. The Basic Engine (BE) recalls the top-\textit{M} most relevant candidate items for the user, where \textit{M} is typically less than two thousand. Our \ourmeth model is deployed on the Real-Time Prediction (RTP) platform to score these top-\textit{M} candidate items. The top-\textit{K} items with the highest scores are then shown to the user. Usually, the $K$ is set to 10-20. In addition, RTP supports deploying multiple models, ensuring fairness in online A/B testing.

\subsubsection{Online A/B Testing Results}
We conduct online A/B testing to evaluate \ourmeth in two online TIR scenarios over a 14-day period, as shown in Table \ref{tab:onlinetest}. 
We choose DEI2N~\cite{xiao2024deep} as the baseline, which performs the second-best in offline tests. 
Results show that \ourmeth consistently outperforms DEI2N in terms of CTR and Conversion Rate (CVR).
Furthermore, since the exposed items can be classified into two categories—those that belong to the same category as the trigger item and those that belong to a different category, CTR can also be segmented into Same-category CTR (S-CTR) and Cross-category CTR (C-CTR). 
\ourmeth shows significant improvements under both S-CTR and C-CTR metrics, which also shows that \ourmeth has better intent representation capabilities. 
From the perspective of cross-category exposure ratio (CCER), \ourmeth can alleviate the problem of over-convergence of intention understanding states and provide users with more diverse and novel recommendations. 
Due to \ourmeth's excellent online performance, we have deployed it in major TIR scenarios on Alibaba.com platform. 

\begin{table}[tbh]
\vspace{-0.7em}
\setlength{\abovecaptionskip}{0.05cm}
  \caption{Online A/B testing results.}
  \label{tab:onlinetest}
  \begin{tabular}{lcccccc}
    \toprule
    Scenario & Model & CTR & CVR & S-CTR & C-CTR & CCER  \\
    \midrule
    \multirow{2}{*}{TIR 1}&  DEI2N & 4.61\% & 3.96\% & 5.01\% & 2.70\% & 41.09\% \\ \cline{2-7}
    &$\textbf{DUIN}$ & $\textbf{4.75\%}$ & $\textbf{4.08\%}$ & $\textbf{5.25\%}$ & $\textbf{2.81\%}$ & $\textbf{46.81\%}$  \\
    \midrule
    \multirow{2}{*}{TIR 2}& DEI2N & 2.73\% & 1.59\%& 2.81\% & 2.92\% & 27.75\% \\ \cline{2-7}
    &$\textbf{DUIN}$ & $\textbf{2.82\%}$ & $\textbf{1.64\%}$ & $\textbf{2.88\%}$ & $\textbf{3.01\%}$ & $\textbf{28.50\%}$ \\
  \bottomrule
  \end{tabular}
\setlength{\belowcaptionskip}{0.9cm}
\vspace{-1.0em}
\end{table}

\subsubsection{Case study}
\begin{figure}[tb]
  \centering
 \setlength{\abovecaptionskip}{0.2cm} 
 \includegraphics[width=1\columnwidth]{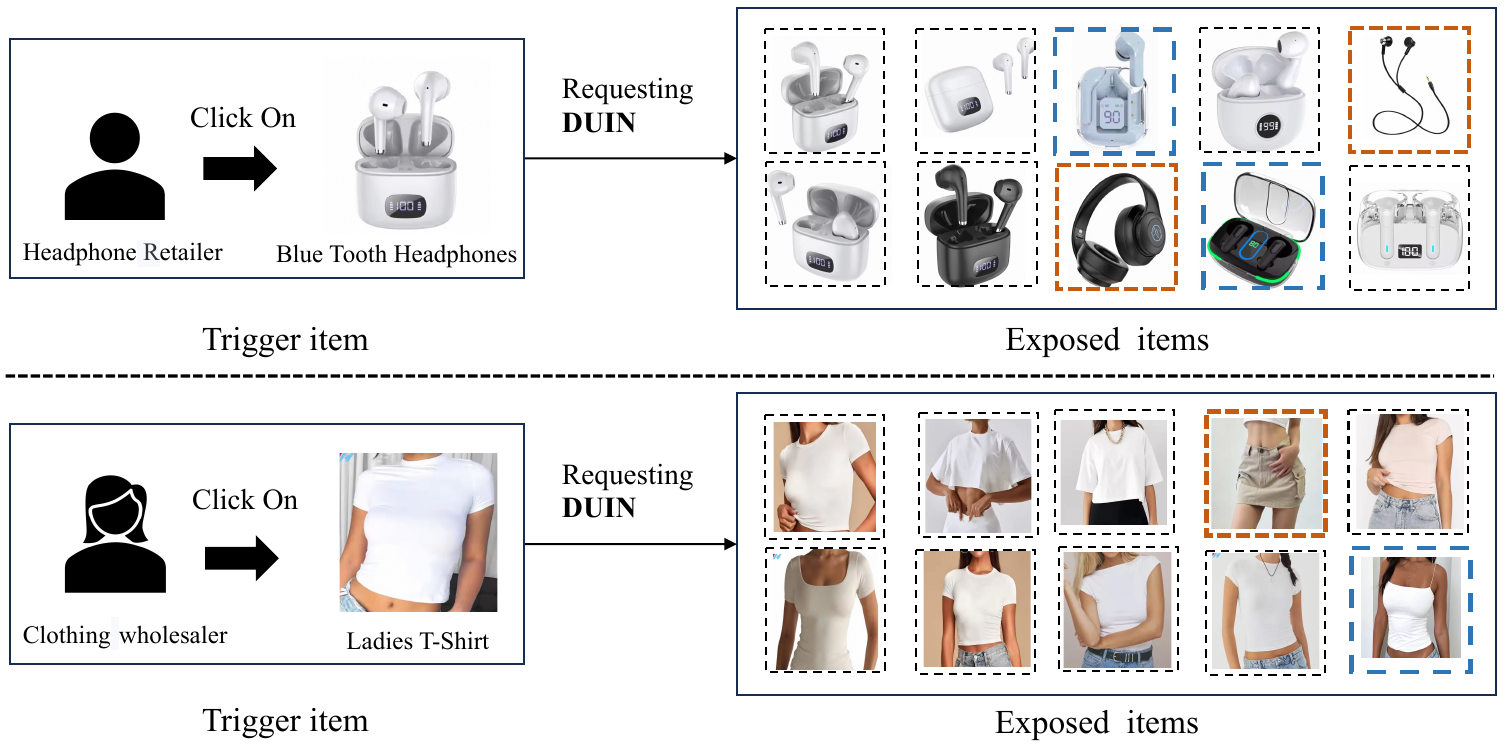}
  \caption{Two real-world cases. Black dashed boxes are similar items, red dashed boxes indicates complementary items, and blue dashed boxes indicates trending items.}
  \label{fig:case}
  \setlength{\belowcaptionskip}{0.5cm}
  \vspace{-0.5cm}
\end{figure}

We illustrate the effectiveness of \ourmeth through two real cases on the Alibaba.com platform. 
As depicted in Figure~\ref{fig:case}, the upper section illustrates that when a user, for instance, a headphone retailer, engages in our TIR scenario. When he/she clicks on the \textit{Blue Bluetooth Headphones}, \ourmeth suggests items that are akin to the trigger item. 
At the same time, \ourmeth suggests trending items and complementary options to the user, like the \textit{Wired Earphone} and \textit{Headset}. 
This shows that \ourmeth not only recommends items similar to the trigger item but also accurately predicts the user's intended scope, offering more precise and comprehensive procurement recommendations. 
The lower section of Figure~\ref{fig:case} shows another real-world scenario. After a clothing wholesaler clicked on the \textit{Ladies T-Shirt}, \ourmeth not only suggests some \textit{Ladies T-Shirts} similar to the trigger item but also recommends \textit{Mini Skirts} that can be easily coordinated with the \textit{Ladies T-Shirt}, as well as \textit{Cotton Tank Tops} that are currently trending on the platform. 
\begin{figure}[tb]
  \centering
 \setlength{\abovecaptionskip}{0.2cm} 
 \includegraphics[width=1\columnwidth]{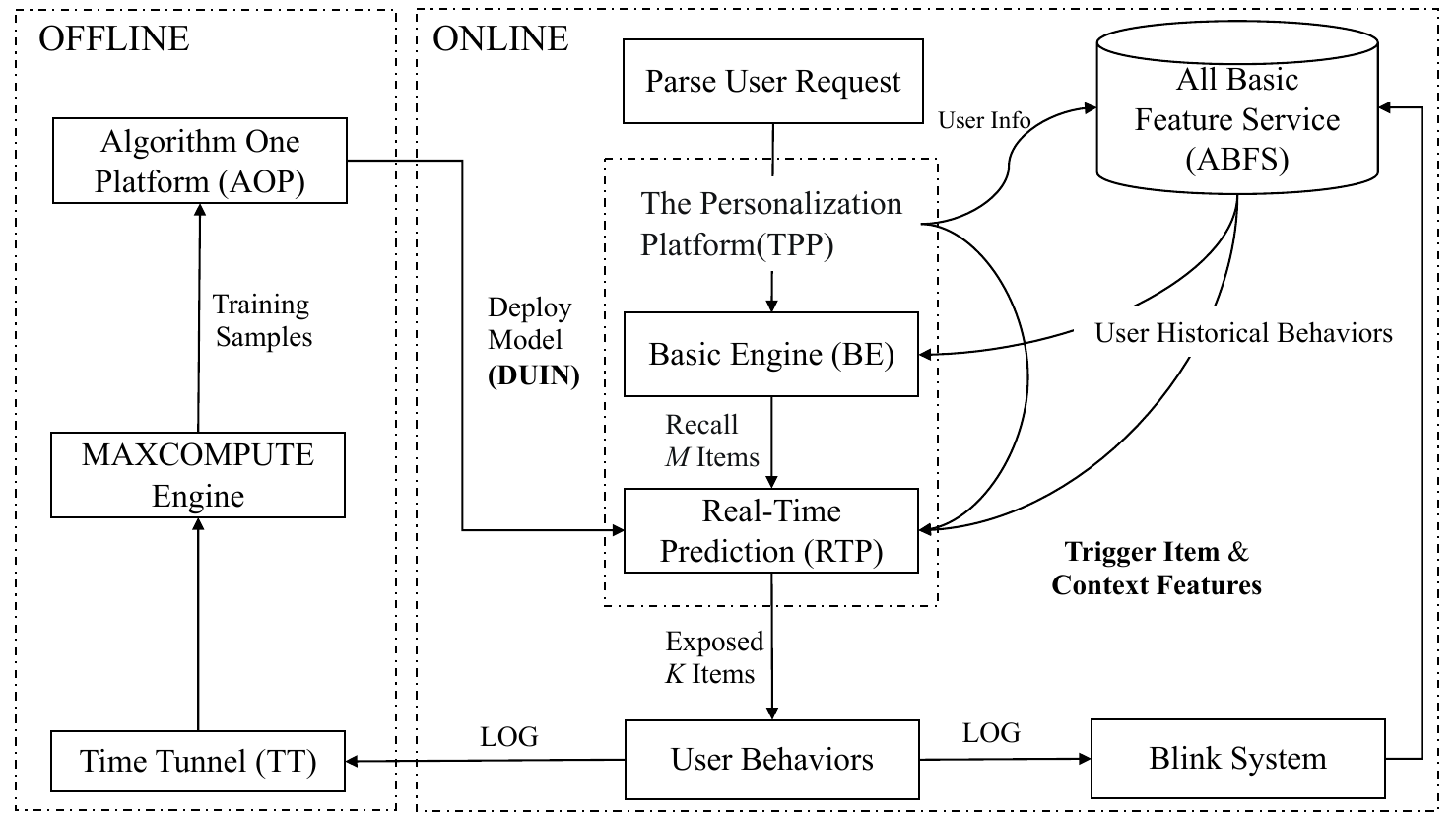}
  \caption{The online serving architecture of \ourmeth.}
  \label{fig:serving}
  \setlength{\belowcaptionskip}{0.5cm}
  \vspace{-0.5cm}
\end{figure}

\section{Conclusion}
In this paper, we introduce a novel TIR model, called Deep Uncertainty Intent Network (\ourmeth), to model both explicit and latent user intent with uncertainty based on user-triggered items. 
The proposed \ourmeth model consists of three modules: i) Explicit Intent Exploit Module, which aims to obtain generalizable and distinguishable intent representations; ii) Latent Intent Explore Module, designed to explore the user’s underlying intent; and iii) Intent Uncertainty Measurement Module, tasked with modeling user intent intensity as a distribution to capture the uncertainty associated with user intent.
Extensive experiments demonstrate that \ourmeth surpasses various representative methods, achieving state-of-the-art performance. Moreover, \ourmeth has been deployed across all TIR scenarios within our Alibaba.com platform, and online A/B testing has conclusively verified the superiority of our approach.

\bibliographystyle{ACM-Reference-Format}
\bibliography{recom_lite}

\end{document}